\documentclass[3p,times,twocolumn]{elsarticle}
\usepackage{amssymb}
\usepackage{graphics}
\usepackage{psfrag}
\usepackage{epsfig}
\begin{document}
\begin{frontmatter}
\title{The proton radius: From a puzzle to precision}
\author[D,E]{Hans-Werner Hammer}
\author[B,J,T]{Ulf-G. Mei{\ss}ner}
\address[D]{Institut f\"ur Kernphysik, Technische Universit\"at
   Darmstadt, 64289 Darmstadt, Germany}
\address[E]{ExtreMe Matter Institute EMMI, GSI Helmholtzzentrum f\"ur
   Schwerionenforschung GmbH, 64291 Darmstadt, Germany}
\address[B]{Helmholtz Institut f\"ur Strahlen- und Kernphysik and Bethe Center
   for Theoretical Physics, Universit\"at Bonn, D-53115 Bonn, Germany}
\address[J]{Institute for Advanced Simulation and Institut f{\"u}r Kernphysik,
            Forschungszentrum J{\"u}lich, D-52425 J{\"u}lich, Germany}
\address[T]{Tbilisi State University, 0186 Tbilisi, Georgia}

\begin{abstract}
We comment on the status and history of the proton charge radius determinations.
\end{abstract}  

\end{frontmatter}

The proton charge radius $r_P$ is a fundamental quantity in particle physics,
as it challenges our understanding of the so successful Standard Model in the
non-perturbative regime of the strong interactions.
It is defined by the slope of the proton charge form factor $G_p^E (t)$
at zero momentum transfer,
\begin{equation}
  r_p^2 = 6 \frac{dG_p^E(t)}{dt}\Big|_{t=0}~,
\end{equation}
with $t$ the invariant four-momentum transfer squared.
The proton charge radius was first indirectly measured in the Nobel prize winning
electron scattering experiments by Hofstadter
et al.~\cite{Hofstadter:1956,Hofstadter:1957wk},
who fitted the form factor data with a dipole form and extracted the
radius from the slope of the dipole. While electron scattering was the
method of choice to refine the measurements of the proton radius in the
decades following these pioneering experiments, the
Lamb shift in electronic hydrogen and muonic hydrogen is also sensitive
to the proton radius~\cite{Karplus:1952}.
%as first suggested by Hans Bethe.
These are electromagnetic bound states of an electron (a muon) with a proton,
where the finite size of the proton leaves a small imprint in the
energy spectrum. Such determinations,
however, require precision experiments and precision theory and thus came
into the  game only later, with a much higher sensitivity to $r_p$
for muonic hydrogen. This is due to the larger muon mass,
$m_\mu/m_e \simeq 200$, so that the corresponding Bohr radius is
smaller and the effect of the proton radius much enhanced.
Most electron scattering experiments gave the so-called {\bf large} radius,
$r_P\simeq 0.88\,$fm, which was
also the value given by CODATA~\cite{Mohr:2008fa}. This was
also consistent with the determination from the electronic
Lamb shift. It then came as a true surprise to most researchers
(but not all, see below) when  the first measurement of the muonic
hydrogen Lamb shift led to the so-called {\bf small} radius,
$r_P= 0.84184(67)\,$fm, differing by 5$\sigma$ from the
CODATA value~\cite{Pohl:2010zza}. The plot further thickened when
high-precision electron-proton  scattering data from the Mainz Microtron
(MAMI) reinforced the large radius~\cite{Bernauer:2010wm}, which was
also consistent with the average value from electronic Lamb shift
measurements, see, e.g.,~\cite{Mohr:2008fa}. Another measurement of muonic
hydrogen, however, supported the small value~\cite{Antognini:1900ns}.
This glaring discrepancy in such a fundamental quantity, which was believed to
be understood since long, became known as the ``proton radius puzzle'',
that featured prominently in many print and online media. For
a review see, e.g.,~\cite{Pohl:2013yb}.

However, while this led to a large number of  publications
scrutinizing the experimental and theoretical approaches, or even questioning
the lepton universality underlying the
Standard Model (cf.~Ref.~\cite{Pohl:2013yb}),
this is not the whole story. Electron scattering
data leading to the proton and neutron charge and magnetic form factors
are best analyzed using dispersion relations, as these embody the general
principles of unitarity, crossing and analyticity. In particular, the
contribution from the closest singularity in the momentum transfer $t$,
the two-pion continuum, can be included in a model-independent
fashion and is of utmost importance for a proper
extraction of the proton radius~\cite{Hohler:1974eq}.
This approach was pioneered and utilized first 
by the Karlsruhe group~\cite{Hohler:1976ax}, and further developed and made
consistent with  symmetries and constraints from Quantum Chromodynamics (QCD)
by the Bonn-Mainz group~\cite{Mergell:1995bf,Hammer:1996kx}.
In fact, the proton radius
puzzle was anticipated in the 2007 paper by Belushkin and the authors
based on a thorough dispersion-theoretical analysis of the
world data base of nucleon
form factor data~\cite{Belushkin:2006qa}. Again, a small radius in the
range $r_p = 0.82 \ldots 0.85\,$fm was found and  it was further shown
that a large proton radius $r_P = 0.88 \ldots 0.90\,$fm could hardly be
accommodated by the form factor data if the constraints from
unitarity and analyticity are taken into account. The
reanalysis of the exquisite MAMI data from 2010 using the same
dispersion-theoretical framework also led to a small
radius of $r_p = 0.84(1)\,$fm~\cite{Lorenz:2012tm}, similar to
all other such analyses before. This was later refined including effects
from the two-photon exchange and performing an improved error analysis,
leading to $r_p = 0.840\; (0.828 - 0.855)\,$fm~\cite{Lorenz:2014yda}.

Still, the situation remained unsatisfactory as $r_p$ from the
electronic Lamb shift was on the large side and there
has been on-going debate about the extraction of the radius from
electron scattering experiments. The situation changed, however,
dramatically when three new experiments on the electronic Lamb shift
\cite{Beyer:2017gug,Fleurbaey:2018fih,Bezginov:2019mdi}, a novel measurement
of electron-proton scattering at unprecedented small momentum transfer~\cite{Jlab},
and another dispersion-theoretical inspired analysis of
electron scattering data~\cite{Alarcon:2018zbz} became available in the
last few years, with the latter one just reinforcing the claims made by the
Bonn-Mainz group since the mid 1990ties. With the exception of the Paris electronic
Lamb shift measurement~\cite{Fleurbaey:2018fih}, all of these new determinations
of $r_p$ consistently
give a small proton radius. Consequently, the newest addition of the CODATA
compilation lists the proton charge radius as $r_p = 0.8414(19)\,$fm~\cite{CODATAnew},
completely consistent with the value from muonic hydrogen and
electron scattering data analysed using dispersion theory.
In light of these results, an improved dispersion-theoretical
analysis including the most recent high-precision determination of the
two-pion continuum contribution~\cite{Hoferichter:2016duk} based on the
tremendously successful Roy-Steiner analysis of pion-nucleon
scattering~\cite{Hoferichter:2015hva} should be performed.
Complementary scattering experiments using muons are also
planned~\cite{Gilman:2013eiv,Dreisbach:2019pkc}.

\begin{table}[t]
\caption{Modern extractions of the proton charge radius from
  the  electronic Lamb shift and electron-proton scattering.}
\begin{center}
\begin{tabular}{|c|c|c|c|}
\hline
$r_p$ [fm] & year &  method & Ref. \\
\hline
0.8335(95) & 2017 & el.  Lamb shift &\cite{Beyer:2017gug}\\
0.877(13)  & 2018 & el. Lamb shift &\cite{Fleurbaey:2018fih}\\
0.833(10)  & 2019 & el. Lamb shift &\cite{Bezginov:2019mdi}\\
0.831(7)(12) & 2019& $e-p$ scattering & \cite{Jlab}\\
\hline
\end{tabular}
\end{center}
\end{table}  

In summary, in view of the new extractions of the proton charge radius
from electronic Lamb shift measurements and very low-energy electron-proton
scattering as well as the on-going activities
to analyze electron scattering data using dispersive methods, we can
now consider the proton radius puzzle solved and look forward to an
increased precision in the determination of this fundamental quantity.

%%%%%%%%%%%%%%%%%%%%%%%%%%%%%%%%%%%%%%%%%%%%%%%%%%%%%%%
\vskip 0.3cm 
\noindent
{\bf Conflict of interest}
    
The authors declare that they have no conflict of interest.
%%%%%%%%%%%%%%%%%%%%%%%%%%%%%%%%%%%%%%%%%%%%%%%%%%%%%%%
%\pagebreak
\vskip 0.3cm 
\noindent
{\bf Acknowledgments}

We thank Hans Str\"oher for pertinent comments and Bastian Kubis
for a careful reading of the manuscript.
This work was supported in part by the Deutsche Forschungsgemeinschaft
(DFG, German Research Foundation) -- Pro\-jekt\-num\-mer 279384907 --
CRC 1245 and the DFG and the NSFC through
funds provided to the Sino-German CRC 110 ``Symmetries and
the Emergence of Structure in QCD''. The work of UGM was also 
supported by the Chinese Academy of Sciences (CAS) President's
International Fellowship Initiative (PIFI) (Grant No. 2018DM0034) and
by VolkswagenStiftung (Grant. No. 93562).
The work of HWH was also supported by the German Federal Ministry of
Education and Research (BMBF) (Grant no. 05P18RDFN1).

%
%%%%%%%%%%%%%%%%%%%%%%%%%%%%%%%%%%%%%%%%%%%%%%%%%%%%%%

\end{document}